# Understanding and Improving Data Repurposing

**Jeffrey Parsons**
Faculty of Business Administration, Memorial University of Newfoundland
St. John's, NL A1B 3X5 Canada {jeffreyp@mun.ca}

**Roman Lukyanenko**
McIntire School of Commerce, University of Virginia
Charlottesville, VA 22908 U.S.A. {romanl@virginia.edu}

**Brad N. Greenwood**
School of Business, George Mason University, 4400 University Drive,
Fairfax, VA 22030 U.S.A. {bgreenwo@gmu.edu}

**Caren B. Cooper**
College of Natural Resources, North Carolina State University
Raleigh, NC 27695 U.S.A.{cbcoope3@ncsu.edu}

**Abstract.** We live in an age of unprecedented opportunities to use existing data for tasks not anticipated when those data were collected, resulting in widespread data repurposing. This commentary defines and maps the scope of data repurposing to highlight its importance for organizations and society and the need to study data repurposing as a frontier of data management. We explain how repurposing differs from original data use and data reuse and then develop a framework for data repurposing consisting of concepts and activities for adapting existing data to new tasks. The framework and its implications are illustrated using two examples of repurposing, one in healthcare and one in citizen science. We conclude by suggesting opportunities for research to better understand data repurposing and enable more effective data repurposing practices.

# 1 Introduction

The expansion of technologies that generate data at scale—including enterprise systems, sensors, health technologies, the Internet of Things, mobile apps, crowdsourcing and citizen science projects, and social media platforms—creates unprecedented opportunities for repurposing data to facilitate analysis and action by managers, policy makers, scientists, and citizens. Data repurposing is a widespread and continuously expanding practice (Goasduff, 2021). It is on the rise in public and corporate settings. Adapting existing data for new uses increasingly supports product and service development, data analytics, healthcare management, marketing and customer insights, journalistic inquiries, and strategy and policy making (Plotnikovs, 2024; Tong & Zuo, 2021). It lies at the core of modern AI, which aims to extract novel patterns from data (Padmanabhan et al., 2022), as exemplified by large language models (LLMs) that fuse many data sources into latent representations from which responses to prompts can be given. Data repurposing also plays an important role in scientific research due to the growing trend of data sharing (Hey et al., 2009). It facilitates computational theory construction, which analyzes data for new theoretical insights (Miranda et al., 2022; Tremblay et al., 2021), and econometric analysis that combines different datasets to reveal relationships between variables of interest (Greene, 2012). The success of these initiatives depends on how data were created, acquired and processed, giving rise to a new area of data management focusing on data repurposing.

We contrast *data management for repurposing* with original (or primary) data management, a traditional area concerned with original data, intentionally designed, collected, stored, and manipulated to support predetermined tasks. This is epitomized by the development of relational databases to support transaction processing (Chua et al., 2022). Original data management commonly uses strict protocols that determine what data are collected, along with standardized structures for processing, storing, and retrieving data. These activities help ensure that data are suitable for defined tasks—the traditional *fitness-for-use* perspective on data quality (Wang and Strong, 1996) — and that data use adheres to legal, ethical, and social standards.

Data management for repurposing differs from traditional data management in important ways derived from the fact that the new uses were not anticipated when the original data were collected. There is often a limited connection between original and new uses, suggesting that a fitness-for-use perspective on data quality needs to be supplemented with a view that considers *use-agnostic* qualities. Data to be repurposed initially will be misaligned with new tasks and, as efforts to align data are undertaken, original data need to be augmented with data from other sources. This involves changing the schema of the original data by adding features that capture entities, attributes or relationships not anticipated in the original design or data collection. In addition, careful attention must be paid to ensure this is done in semantically meaningful, operationally useful, and ethically appropriate ways. This, in turn, requires that data be accessible, its semantics transparent, and that it be rich enough to accommodate unanticipated uses.

Because of its unique characteristics, data repurposing brings new data management challenges. However, research has not kept pace with the opportunities arising from repurposing. Most work has been rather narrow – in the context of scientific data reuse (which may or may not involve actual repurposing) or on technical topics such as semantic data integration. The constituent concepts and activities have been developed independently across heterogeneous contexts and lack a strong foundation and common understanding. This has been acknowledged by practitioners, who have made calls to "rethink how data management must be conducted" (Strengholt, 2020, p. 1). Concerns have been raised that many data management activities in repurposing, such as how to evaluate dataset quality (Sadiq et al., 2022), are "invisible" due to poor understanding of them (Parmiggiani et al., 2022). The lack of established practices increases the costs of locating, evaluating and preparing data (Woodall, 2017). Finally, there is no consensus on what constitutes repurposing, how it differs from other related activities (e.g., data reuse) and from original data management, and how to repurpose data effectively and responsibly.

This commentary develops a framework and research agenda for data repurposing across scholarly and practitioner communities. We define repurposing as practices that augment the schema of existing data with additional entities or attributes in a manner that goes beyond what was designed to satisfy the original purpose. This is contrasted with data reuse, which can be accommodated using an existing schema without any changes. The framework articulates the concepts and activities involved in repurposing, the key data objects involved, the actors that perform the activities, the enablers and constraints of repurposing, and its antecedents and consequences. We illustrate the framework and its value by applying it to two repurposing uses cases, one in public policy/healthcare and the other in citizen science. Finally, we use the framework to propose a data management research agenda to better understand data repurposing and to improve repurposing practices.

## 2   Research on Data Repurposing

While data management research has predominantly focused on original data (e.g., Chua et al., 2022; DAMA et al., 2024), a notable body of work studies the challenges and opportunities of combining heterogeneous data and using them in different contexts and for new tasks.

First, much thinking about using existing data for new tasks has been in the context of open science, which supports sharing data to answer new questions, enable meta-analysis (Briney, 2015), and accelerate discovery (e.g., GenBank, www.ncbi.nlm.nih.gov/genbank, an open archive for DNA sequence data). The impetus for sharing comes from funding agencies that increasingly require applicants to commit to open data sharing and provide a corresponding data management plan.[1] Open data raises numerous challenges along ethical, technical, and policy dimensions. For example, data sharing requires considerations of consent (a challenging issue given the impossibility of envisioning and articulating all possible ways data can be later used by others), anonymization and other privacy protections, and legally binding license agreements (Briney, 2015; Law, 2006). Other challenges involve determining how to assure data quality, standardize data, and describe data to facilitate interpretation by others (Bono et al., 2023).

To address these challenges, researchers and agencies have proposed best practices for scientific data management. For example, the Findable, Accessible, Interoperable, and Reusable (or FAIR) principles encourage scientific data sharing through provision of rich meta-data, free and open data access protocols, shared and formalized knowledge representations, and clear licensing (Wilkinson et al., 2016). Yet, despite the scientific focus of these efforts, they are generally not guided by theory and do not articulate the fundamental characteristics of repurposed data. Furthermore, scientific data management has several unique characteristics (e.g., emphasis on open data sharing) that are absent in other contexts, and prior work in this area does not distinguish between data reuse and repurposing.

---

[1] For example, European Union, https://op.europa.eu/en/publication-detail/-/publication/7769a148-f1f6-11e8-9982-01aa75ed71a1/language-en/format-PDF/source-80611283

Second, a large body of technical research relevant to repurposing deals with the challenge of combining data from multiple sources. This work emerged from the growth of independent databases and the recognition that integrating multiple sources is necessary to answer certain types of queries or to deal with the need to link related data due to mergers and other corporate restructuring. Research addressing the need to integrate data falls under the scope of semantic data integration and database interoperability (Batini et al., 1986; Park & Ram, 2004).

Two main approaches to the data integration/interoperability challenge have been studied. One comprises techniques to combine multiple partial views of data requirements to develop a global schema prior to implementing a single database (Batini et al., 1986; Parsons, 2002; Spaccapietra & Parent, 1994). In that work, the design of a global schema aims to reconcile independent views and provide a model of data suited for a set of anticipated uses. Most work done on this topic has focused on database design and use in organizational settings.

The second approach deals with combining data from existing independent databases (Castano & De Antonellis, 2001; Litwin et al., 1990), where a key challenge is reconciling syntactic and semantic differences in data sources to answer specific queries (Friedman et al., 1999). Various strategies have been used to address this challenge, including schema mapping (Miller et al., 2000; Park & Ram, 2004), federated databases (Sheth & Larson, 1990), and query translation (Friedman et al., 1999). Again, most work on this topic has dealt with traditional organizational databases. In addition, although this work considers querying across multiple sources, little attention has been paid to the nature of data and its potential for repurposing.

The problems of data integration and semantic reconciliation have amplified as the number and diversity of data sources—no longer limited to relational databases—has grown. This has resulted in a stream of research aimed at data integration for the Semantic Web (Langegger et al., 2008) and querying multimodal data sources (Xiong et al., 2025).

Notwithstanding considerable progress, data integration work has focused on structured tabular data and semantically matching entity types and attributes across multiple sources. This provides a good technical foundation for semantic integration, but does not address the characteristics of data that support unanticipated uses or the process of repurposing data.

Another stream of technical work focuses on the tools, infrastructure, and methods of managing big data, often with the aim of supporting data reuse and repurposing. Of particular interest are highly scalable storage technologies, such as the Apache Hadoop, NoSQL databases, data lakes, and federated data meshes (Gorelik, 2019; Inmon & Srivastava, 2023; Plotnikovs, 2024). Practitioners are frequently the key drivers of new technologies, as the data repurposing challenges they experience appear to have no solutions in academia or in the data management body of knowledge.

Third, a notable emerging area of research is on the quality of secondary data. This work has been especially active in the context of data analytics, where metrics and methods of assessing data quality for reuse have been developed (Kenett & Shmueli, 2016; Zhang et al., 2019). Nevertheless, important questions remain. For example, a common assumption has been the existence of a dataset that can be evaluated for a new purpose, leaving open the question of whether and how to assess the repurposability of a dataset prior to the identification of new tasks. As Sadiq et al. (2022) lament: "[t]he body of knowledge on how to evaluate the quality of datasets that exhibit characteristics typical of re-purposed data is critically lacking."

Finally, a growing body of research investigates the impact of evolving data use trends on individuals, organizations, and society (Davenport & Harris, 2017; Yoo et al., 2024). This work is supported by conceptualizing data as a valued asset (Aaltonen et al., 2021; Baskerville et al., 2020). However, these studies do not adopt a data management perspective on repurposing (hence, issues such as data quality and data modeling are not addressed). Instead, the goal is to understand the evolving nature of data and appraise the economic and societal significance of "data making" (Alaimo & Kallinikos, 2024). A review of perspectives on digital data concludes that recent data practices, including reuse, are ill-understood (Aaltonen et al., 2023). Table 1 summarizes key themes of prior research on data reuse and repurposing.

| Table 1: Themes of Prior Research on Data Reuse and Repurposing | | | |
|---|---|---|---|
| **Theme** | **Research Focus** | **Major Outcomes** | **Open Questions** |
| Reuse of scientific data | Enabling data sharing among researchers to enhance scientific progress | • Guidelines for data sharing<br>• FAIR principles | • What characteristics of data and the environment in which is it used contribute to its repurposability? |
| Data Integration | Interoperability of data from multiple independent sources | • Schema (view) integration techniques<br>• Database integration techniques<br>• Query rewriting | • How to extend techniques to heterogeneous and multimodal data environments?<br>• How to extract schemas from semi-structured and unstructured data?<br>• How to combine structured data with semi-structured and unstructured data sources? |
| Infrastructure for big data management | Scalable storage technology and related data access | • Data lakes, data meshes<br>• ML-friendly data access | • What are the benefits and limitations of technological solutions to repurposing data in semantically appropriate ways? |
| Data quality for reuse | Evaluation of data quality for reuse at time of reuse | • Methods for assessing data quality<br>• Metrics for data quality | • Can the repurposability of a dataset be assessed prior to reusing the data for a particular task?<br>• How can we assess (and bridge) the gap between available data and data needed for a new purpose? |
| Implications of data reuse and repurposing | Impact on individuals, organizations, and society | • Conceptualizing data as a valued asset | • Can we fully consider the implications of repurposing without a comprehensive framework of data repurposing? |

In sum, a comprehensive and theoretical understanding of repurposing is absent, resulting in *ad hoc* repurposing practices. An impediment to studying repurposing is the proliferation of concepts related to using data for new tasks, referred to by different and often not synonymous terms across communities (e.g. data use, data reuse, data repurposing). Likewise, the data being used for new purposes is referred to using different names (e.g. *secondary*, *tertiary*, *archival*, *administrative*, *third-party, data product, data commodity*), each having a distinct emphasis that may or may not involve repurposing.[2] It is unclear what distinguishes repurposing from original data use or data reuse and therefore what, if any, practices are unique to repurposing. Little attention has been paid to how to manage original data with repurposing in mind or how to assess the gap between the obtainable data and the ideal data for the new task. Next, we address this by proposing a general framework for data repurposing.

## 3  Data Repurposing Framework

Adapting data for new tasks is a complex and challenging process. Recognizing these challenges, academia and industry have called for rethinking traditional approaches to data management (Aaltonen et al., 2023; Strengholt, 2020). We develop a framework that articulates the key components of data repurposing to support and drive further research on the topic, and to guide practitioners on important considerations when repurposing data.

### 3.1  Understanding Data Repurposing

A prerequisite for understanding data management for repurposing is defining what data repurposing is and how it differs from original data use. Fundamentally, data are complex symbols (Alaimo & Kallinikos, 2022; Lukyanenko & Weber, 2022). A symbol is a non-iconic sign that represents something and is interpreted by convention, rather than by similarity to its referent. Symbols are "*means-objects*", serving some purpose (Morris, 1946, p. 368). Accordingly, data always have an original purpose. Even trace data (e.g., sensor data, server logs) are products of the design of the system that generates these logs.[3] As the name implies, repurposing involves using something for a purpose other than the original one.

We now establish the nature of data repurposing and distinguish it from original data use and data reuse (Figure 1). Original data is created with a *purpose* to facilitate some *task(s)* – to support an action or answer a question. The purpose of data is encapsulated in a *schema* – an explicit or implicit conceptual representation that provides the blueprint for the structure and properties of data such that the data can satisfy that purpose. These representations can range from formal conceptual data models (such as an entity-relationship diagram) to informal representations (such as textual narratives that describe specific use cases for data).

The schema converts the task requirements into data abstractions, such as entity types, attributes, and cardinalities. It also embeds assumptions, values and beliefs about what must be captured and what can be ignored based on the intended use. Guided by the schema, data can be captured, stored, and used to facilitate the desired task. Therefore, the data schema supports the original purpose and the data conform to the schema. When made explicit, the schema may accompany the data as a diagram, metadata, or a data dictionary.

---

[2] For example, secondary data is especially popular in research (e.g., Law, 2006), referring to data used by someone other than its creator. As we show later, repurposing can be performed by the same person, and data transformed during repurposing may be further repurposed (and thereby become tertiary data).
[3] For example, server logs are frequently used for security and auditing, which informs software developers of the ways to provide records of the server activities. As such trace data are often repurposed, for example, to understand consumer behavior (Berente et al., 2019), the way it is generated must be taken into consideration. This might involve actions such as removing potentially identifying information to preserve anonymity.

Consequently, we define *original data use* as planned data use that led to the design of the original data schema. Therefore, original data use involves anticipated querying and manipulation (e.g., visualization, statistical aggregation) over data that conform to the original data schema. Some data manipulations (e.g., new ways to visualize) may not be anticipated at the time when data were originally modeled and collected and may even be performed by different people in a new setting. However, as long as they can be performed with the original schema, they do not constitute repurposing. We refer to tasks that involve asking new questions of the data that conform to the original schema as *data reuse* (see New Task 1 in Figure 1).

In contrast, we define *data repurposing* as data use performed after making one or more additions (e.g., new attributes, new entity types) to the schema describing the original data. Such changes imply that augmenting the original schema is part of the data transformation required to support the new task (otherwise the existing schema would be sufficient, making it data reuse). Data repurposing may start either by formulating a *task*, which drives the process to obtain suitable data (task-driven repurposing), or by realizing that existing data may be augmented to support tasks other than those for which they were collected (data-driven repurposing).

Unlike schema integration procedures used in data integration, suitable additional data sources must be discovered, often via the Internet. Furthermore, the schema might need to be inferred from unstructured sources. Data-task alignment involves elucidating an ideal schema for the task, finding candidate data, and evaluating the match of candidate data to the task based on the schema of the available data and the ideal schema for the task. Once potentially suitable data is found, transforming the original data to align it with the ideal schema can proceed using data integration techniques (Batini et al., 1986). However, as repurposing may involve combining data from disparate domains and answering questions not anticipated when original data were designed, the impact of integration becomes evident only after determining whether the transformed data can support the new task. If unsuccessful, the data may be discarded or the task modified.

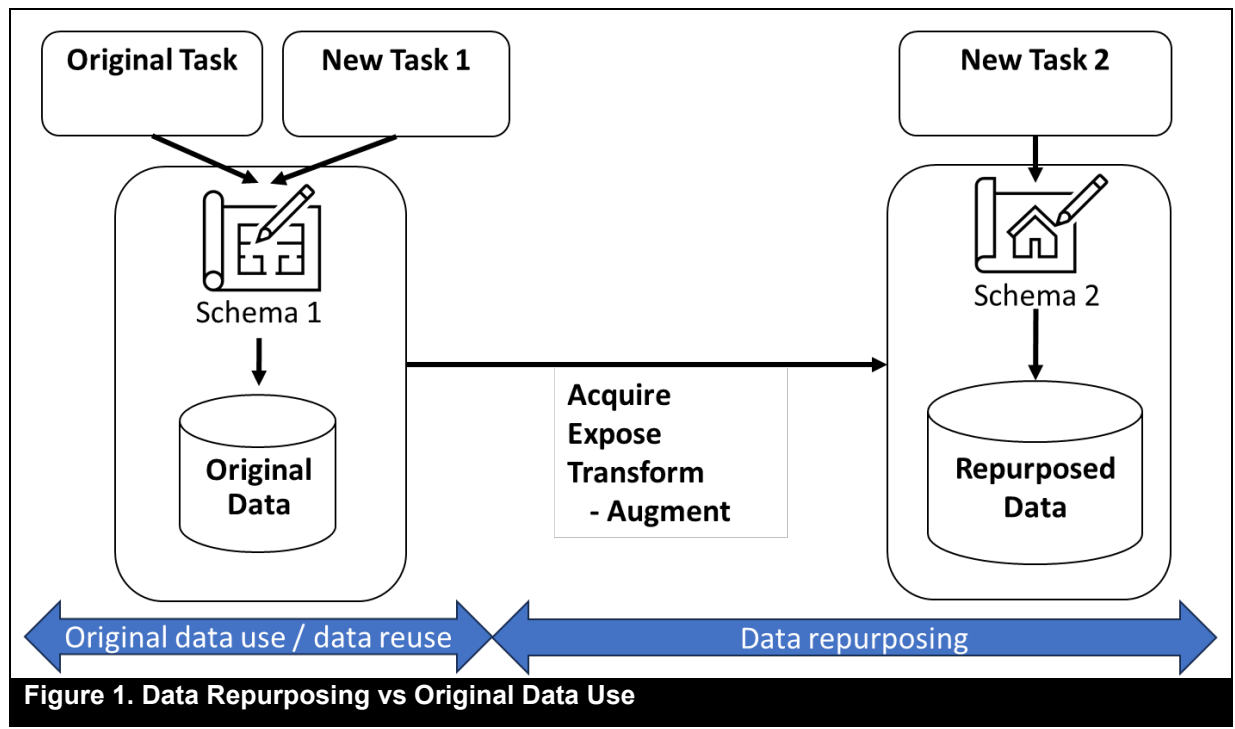

**Figure 1. Data Repurposing vs Original Data Use**

The extent of data-task alignment depends on the similarity of the ideal schema to the original one, making it possible to establish the limits of repurposing. Repurposing is possible as long as there is an overlap in attributes between the original schema (Schema 1 in Figure 1) and the ideal schema for the new task (Schema 2 in Figure 1). If there is no schema overlap, the existing data cannot support the new task. In that case, new data acquisition is required.

Critically, knowledge of the original context is important for understanding how the data were shaped, what was purposefully or inadvertently left out, and how they were curated. If the schema of the original data was explicit and detailed enough to capture this knowledge, it can be used to better understand the original context of creation. However, the schema is often missing, contradictory, or incomplete, creating a challenge to infer the original context from the data.

At the same time, recontextualizing data (see Aaltonen et al., 2021; Alaimo & Kallinikos, 2024) does not encompass all repurposing concerns. Indeed, repurposing may occur in the same setting and by the same people as the original data creation. While such a situation makes certain repurposing tasks easier (e.g., understanding what the original data means), challenges remain. For example, Plotnikovs (2024) shows that even when the same data teams repurpose their data, a significant challenge is to determine whether the original data are suitable (e.g., granular enough, of sufficient accuracy) for the new tasks. We consider these issues next.

## 3.2 Data Management Activities for Repurposing

Figure 2 outlines the Data Repurposing Framework, comprised of the objects, activities, actors, and environment of repurposing. These concepts apply in all repurposing settings, but there will be variations in how they are expressed depending on the setting. For example, in corporate settings, outcomes such as profitability or operational efficiency are core goals of repurposing. In research repurposing, goals such as discovery or knowledge creation might be the focus. In addition, the environment (enablers and constraints) will differ across repurposing settings, depending on factors such as resources or the regulatory environment surrounding data repurposing (e.g., sensitive patient files vs. public archival data).

**Original data management**. While data always has an original purpose, it also has the potential to be used in different ways. Indeed, any artifact is characterized by *duality of purpose*—serving a specific (original) need, but capable of supporting additional uses as users discover new ways to leverage existing features (e.g., Germonprez et al., 2007). Adapting this to the data context suggests considering the impact of original data management, as it develops *original schema* and results in the **original data**.

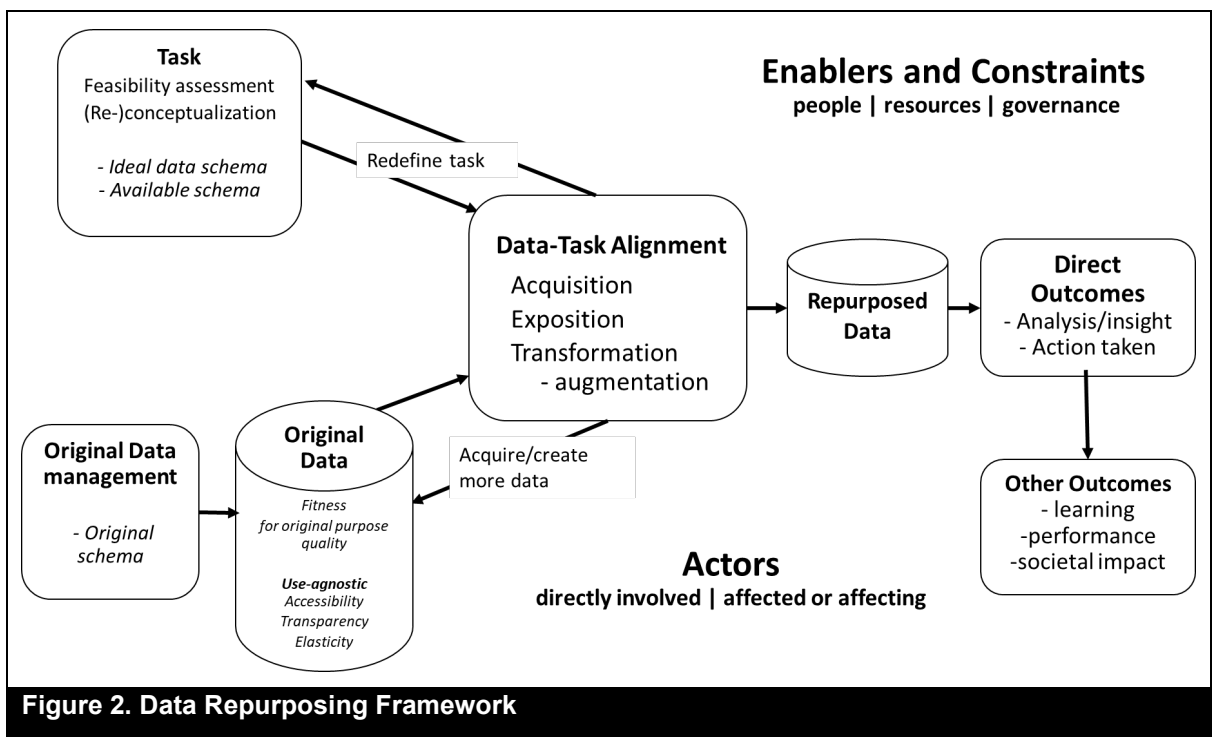

**Figure 2. Data Repurposing Framework**

When repurposing data, original data management considerations remain relevant, although they do not apply to repurposing in the same way as for original data. Chua et al.'s (2022) data management framework articulates the scope of original data management as the "activities and methods to conceptualize, collect, curate, consume and control data to support insight, analysis, and action" (Chua et al., 2022, p. 1). For example, how original data are conceptualized affects their form and structure (Burton-Jones & Weber, 2014). Thus, as part of conceptualization, a data schema may be developed to specify predefined categories, as is done with entity-relationship diagrams. Conceptualizing the same person as a customer vs. store visitor results in different schemas, which in turn affects what kinds of data (e.g., attributes) are stored about this person and how the data can be repurposed. However, while original data management focuses on ensuring the data is suitable for the original task (Chua et al., 2022; Wang & Strong, 1996), little consideration is given to how this focus affects its repurposing potential.

A primary concern of original data management is the ***fitness-for-purpose*** of the original data. Among other concerns, this involves ensuring original data is accurate and complete enough for the original task (Wang & Strong, 1996). However, fitness-for-purpose does not apply in the same way for data being repurposed. For example, data that are accurate and complete enough for the original purpose might not be so for the new purpose. Moreover, as data requirements for unknown tasks cannot be predicted, fitness-for-purpose cannot be determined until a new task is specified. In this case, a relevant property of data is its ability to accommodate potentially unknown future purposes, leading to new data management considerations.

In addition to fitness-for-(original)-purpose, original data have inherent ***use-agnostic*** properties – *accessibility*, *transparency*, and *elasticity* – that enable or inhibit repurposing. These properties directly influence the extent of challenges when repurposing data: finding relevant data, evaluating the suitability of data, and tailoring data to suit the new task. The levels of these properties are shaped by original design decisions but can be altered after data are collected. High levels of these properties help ensure that data can be used across a spectrum of tasks.

*Accessibility* is the extent to which data are discoverable, searchable, and available to support new tasks. For example, text is directly searchable, unlike videos, images or sound, which presently require additional data (e.g., markup tags and descriptions) or advanced processing (e.g., using computer vision techniques to extract visual and audio patterns). Hence, text format is more accessible than video or audio formats. Accessibility is affected by the systems that store and interact with data, as well as by factors such as organizational policies, sharing norms, and legal obligations.

*Transparency* is the extent to which it is clear what the data represent; that is, what aspects (states) of reality the data map to and how the data are to be interpreted. Original data has inherent transparency because the clarity and unambiguity of the mapping is rooted in the fact that data consumers or their proxies commonly produce (or verify) the original data schema and hence, shape the subsequent data management activities (Lee et al., 2006). However, a user who is unaware of the context of data creation might be ignorant of factors such as the motivation for collecting the original data, potential biases in the data, or the capabilities of the data recording instrument. As Alaimo and Kallinikos (2024, p. 66) note, "[a]s data travel across contexts and are reused [or repurposed], the predilections, assumptions, and design choices on the basis of which they are initially produced become opaque and fade into the background." Repurposing data that lacks transparency can be counterproductive and harmful.

Accessibility and transparency of data are necessary for repurposing. For example, the FAIR principles (Wilkinson et al., 2016) consider these two issues. Similarly, studies on data making and data commodities acknowledge the challenge of "interpreting data's meaning" in a new context (Aaltonen et al., 2021; Alaimo & Kallinikos, 2024). What is not explicit in existing discussions of repurposing is the extent to which original data (even if accessible and transparent) is *amenable* for repurposing. Evidence from practice suggests this is a significant challenge, often leading to task redefinition (addressed later), and to harmful outcomes if misaligned data is used for new tasks (e.g., Plotnikovs, 2024).

*Elasticity* is the extent to which original data can be used to support repurposing tasks. The more elastic the data, the more they can be metaphorically stretched to fit multiple schemas to support tasks beyond the one(s) for which they were originally designed.

Consistent with our conceptualization of repurposing (see Figure 1), to be elastic (and amenable for repurposing) original data must be able to conform to another schema; that is, the data should be at least somewhat independent of, or separable from, the original data schema. As a schema is a conceptual blueprint of data, we define *conceptual data independence* as the extent to which data can be decoupled from its original schema and, after some transformations, sufficiently aligned with a new schema to satisfy the demands of a new task. Repurposing thus extends the theoretical notion of data independence (Codd, 1970; Hellerstein, 2003; Parsons & Wand, 2000), to which it also contributes the concept of elasticity. Originally, data independence focused on physical and logical independence (Codd, 1970). However, repurposing leads naturally to broadening the concept to also include data-schema, or conceptual, independence (Parsons & Wand, 2000).

We suggest several antecedents of elasticity. First, due to the way data are presented, it may be difficult for data consumers to recognize the match between the original data and the ideal schema for a new task. For example, data may present a psychological anchor from which decision-makers may not be able to sufficiently adjust their view to see another way to use the data. Research has shown that different data structures produce different anchoring effects (Lukyanenko, Parsons, Wiersma, et al., 2019). Evidence from practice suggests how difficult it is for one data team to view data in a different way because of the way they are used to thinking about it (e.g., as customers vs company partners, see Plotnikovs, 2024). Recognizing anchoring and other psychological biases can help future users view data from a different perspective.

Second, the way data are collected affects elasticity. For example, highly granular data (representing more precise, atomic facts about the world) can be more easily aggregated for future uses than coarse data. The latter may require augmentation with more granular data to meet the needs of a new task. Thus, fine-grained data is inherently more elastic than coarse data.

Third, the way data are stored affects elasticity. To illustrate, consider a relational database consisting of tables in third normal form. Assume, one normalized table, designed to eliminate data insertion, update, and deletion anomalies, is shared for repurposing.[4] To repurpose the data, combining it with other data (via foreign keys) may be required, which depends on knowing where to find the related tables and their accessibility. A more elastic approach would be to design a not "too heavily normalized" table (Strengholt, 2020, p. 171). More generally, data storage is more elastic if it does not require additional information or technology to use the data.

Finally, the way data are governed affects elasticity. Thus, elasticity can be degraded due to low original quality of data. If original data is inaccurate or has missing values, it may create insurmountable challenges for repurposing. Worse still, quality issues may compound, as a minor problem for original use may result in severe misalignment when data are used for a new task. Thus, strong data governance is important to ensure data are amenable for repurposing.

**Task**. *Repurposing feasibility assessment.* The process of repurposing often starts with a task and a belief that existing data might support that task. Still, not all tasks should be undertaken by repurposing data. An important determination is whether using existing data or obtaining new data is better suited for the task at hand. This involves considering the direct and opportunity costs and constraints (e.g., legal, ethical) for each option. The better the understanding of the task, the easier it is to decide which option is most suitable.

---

[4] For example, US Social Security Administration shares *a single table* of births in the US: https://catalog.data.gov/dataset/baby-names-from-social-security-card-applications-national-data.

*Task (re)conceptualization*. When a task is specified, the data needed for the task can be conceptualized, resulting in an *ideal data schema*—a standard against which available or obtainable data can be evaluated. Typically, available data will not match this schema exactly. The degree of match between ideal schema and the available data affects how repurposing can proceed. Given the initial conceptualization of the ideal data for a task and attempts to repurpose available data, it may become clear that the original task needs to be reconsidered. For example, a hospital may wish to use online physician ratings as a proxy for quality of care. However, after identifying data quality issues in online reviews (e.g., selection biases), the task may be redefined to one of understanding why some patients leave online reviews while others do not. Hence, for successful repurposing a match needs to be established between the task and the obtainable data. Still, it is challenging to assess this match without the data-task alignment efforts we consider next.

**Data-task alignment.** Unlike original data provisioned to fit a predefined task, data for a new purpose might not be readily available. In that case, the challenge is to discover and access one or more existing datasets with characteristics matching the task. *Data-task alignment* is the extent to which given data can be used to perform a specific task, or the extent to which a task can be shaped to be performed with available or obtainable data. It can also be considered as the process of achieving a level of alignment. Because the original data were collected for one task, its schema does not match the ideal schema for the new task. Consequently, effort may be needed to align data and task via further data acquisition, exposition, and transformation. Moreover, in aligning data for a new task it is not unusual to redefine the original task in light of preliminary efforts to acquire, expose, transform, and augment data. Such iterative redefinition of the task is an important part of realizing the potential of repurposing data.

*Acquisition*. In task-driven repurposing, once there is a sense of the ideal data schema, the next challenge is identifying the most relevant available data for the task. If the process is data-driven, the initial dataset needs to be augmented with data from other sources to modify its schema to address missing features based on the initial conceptualization of the new task.

Acquisition is challenging given the explosive growth of data sources that lack clear documentation, which can make it difficult to find relevant data. What is more, before data can be repurposed, alignment barriers (i.e., transparency and elasticity) must be overcome. Ensuring that, among the candidate data sets, the one with the best task alignment is selected compounds the difficulty of data acquisition. Acquiring the wrong data can be costly, including the effort required to align the data with the task, potentially paying for access to the data, and the opportunity cost of not using better data.

*Exposition*. Before data can be repurposed, it is critical to have transparency about its meaning. As understood in the data management literature, using data appropriately is contingent on understanding the *context* in which it was created (Shankaranarayanan & Blake, 2017; Strong et al., 1997; Watts et al., 2009). However, context is often vague and ill-defined. To support data repurposing, we formalize *context* as knowledge of the original data management activities to "conceptualize, collect, curate, consume and control data" (Chua et al., 2022, p. 1).

Sufficient exposition of the original context is essential for safe and responsible repurposing. For example, how data was conceptualized can be critical in identifying misinformation and assessing its completeness (Kim & Dennis, 2019; Tong & Zuo, 2021). Similarly, transparency in how data are stored and processed can help in determining if they were altered after being collected. Alterations are often needed to correct original errors or remove sensitive information (e.g., Menon et al., 2005). Still, problems can arise when users are unaware of the alterations or do not understand why they were made. Finally, to ensure safe and appropriate repurposing, the controls involved in original data management must be known. Ensuring that the original data adhered to legal, ethical, and other considerations (e.g., sustainability) is critical. In repurposing, obtaining high levels of transparency presents a different challenge from that in original data management because of the separation between the contexts of creation and use.

Because original data use and reuse commonly allow data users to communicate with those who engaged in data modeling and collection, it can be done without explicitly documenting every aspect of data. As a result, some knowledge is driven by shared understanding of factors such as why data is needed and who created it. This is epitomized by agile development (Johnson, 2020), which is characterized by continuous and close interaction between users and software developers such that systems, including data structures and data insights (in the case of agile analytics projects), emerge through intense collaboration. As a result, documentation that is generally seen as excessive and wasteful is minimized.

In data repurposing, there is a natural distance between data production and use. As a result, even if data analytics is done using agile methods, one cannot rely on shared understanding or tacit assumptions about why data was created and how it ought to be used (e.g., to ensure semantically meaningful and appropriate use). Consequently, these assumptions need to be discovered and clearly elucidated, a process not needed for original data management. Due to the proximity of production and use contexts, exposition as a data management concern may be entirely absent in original data management. In contrast, exposition challenges can present insurmountable barriers to repurposing data.

*Transformation.* If candidate data are sufficiently exposed, various approaches can be taken to align them with the requirements of the new task. We define *transformations* as actions to increase alignment between the task and data. Transformations for repurposing invariably include data augmentation, as a new schema requires bringing in new data to satisfy new schema requirements. When the schemas of the original and new data are available, augmentation can be done using schema integration techniques (e.g., Batini et al., 1986). Transformations also commonly involve actions over the original and new data such as filtering, debiasing, pivoting and granularity alignment. For example, to investigate the impact of online classifieds that facilitate sexual encounters on the rate of sexually-transmitted diseases, information about the regional rollout of such ads must be augmented with health data on the incidence of those diseases (Chan & Ghose, 2014). Thus, the schema must be expanded to include the number of cases in a region over time. The locational and temporal data provide the link (i.e., shared attributes between the original and new schemas) needed to combine the two data sets. In some cases, the necessary data are not available from any existing sources. When this happens, it might be possible to collect additional data, for example, via crowdsourcing.

**Repurposing Outcomes**. Following data-task alignment, a new data set is generated. This *repurposed data* adheres to the schema changes needed to ensure alignment and contains objects such as documentation on transformations and any notes or observations related to the meaning of data. The data can be again repurposed, resulting in the creation of data commodities such as credit scores or company indices (Aaltonen et al., 2021) that can be bought or rented and further repurposed by combining with other data to draw additional insights.

Repurposing *per se* concludes with the completion, abandonment, or redefinition of the task following the analysis of the data. We conceptualize *direct repurposing outcomes* in terms of the degree of completion of the task. Repurposing can also produce *indirect outcomes*, including expanded market share and improved competitiveness, better engagement and retention of customers, improved operational efficiency, the development of new products, services or technologies, and the generation of new knowledge (Freeman, 2010).

While there are direct and indirect outcomes for original use, reuse, and repurposing, there are several differences. One is the unpredictability of the repurposing outcomes, for example, with respect to privacy, security, reputation, consent, or intellectual property. The range of possibilities of unanticipated outcomes is broader when repurposing data because it is very difficult to fully anticipate the range of potential schema changes. Another difference is in uneven competitive advantage resulting from repurposing. Access to repurposing opportunities varies. In particular, data-rich organizations may enjoy a competitive edge due to their ability to combine proprietary data in new ways (e.g., Google search traffic and mapping data to recommend location-based services) (Krämer & Shekhar, 2025). These benefits are impossible to achieve for those lacking rich proprietary data, limiting options for data augmentation to those based on publicly available data, which, being available to everyone, has few barriers to use. As a result, repurposing may bestow a unique competitive edge to some market actors, resulting in a new kind of "economic moat" (Boyd, 2005).

Repurposing can have both positive and negative economic and social impacts. According to the European Union's Open Data Initiative, open data sharing stands to increase the EU's GDP, create new jobs, and usher in substantial cost savings in both the public and private sector.[5] Estimates put the annual global economic value of "enhanced use of open data" at $3 trillion (Chui et al., 2014, p. 1). Still, the success of data repurposing needs to be considered against its costs, including providing the requisite skills, infrastructure, and training, as well as the potentially detrimental consequences for individuals (e.g., erosion of privacy) and society. For example, an overlooked consequence of rampant data manipulation is the environmental impact of handling data at scale. Technologies such as machine learning, especially training large language models, have a growing environmental footprint (Wu et al., 2022).

**Actors of Repurposing.** Repurposing is a complex and ubiquitous activity that affects a wide range of people, both in their work and personal capacities. Furthermore, with automation many tasks are now being performed by artificial intelligence. Thus, there are people and organizations who may not perform a repurposing activity but are affected by it. Often, their needs may not be apparent, and could be neglected both by those conducting repurposing, as well as by data management researchers. Table 2 outlines key actors, both direct and indirect, capturing their roles and key challenges, which we also suggest as research opportunities.

| Table 2: Challenges for Actors Involved in Data Repurposing | | |
|---|---|---|
| **Activity** | **Directly Performing Activities** | **Affected or Affecting** |
| **Original data management** | **Data designers:** How to manage data to satisfy original tasks, while making it more elastic? <br> **IT personnel / support:** How to create infrastructure to make data more accessible, transparent, elastic? | **Data subjects:** How is my data used? Can/should it be repurposed? For any purpose or some? Rescind / update consent? |

---

[5] https://data.europa.eu/en/publications/datastories/economic-benefits-open-data

| | | |
|---|---|---|
| | **Managers:** How to foster organizational data management practices to accommodate new uses of data? | **Regulators:** How to facilitate repurposing while safeguarding privacy? How to promote data-driven society? How to promote / incentivize data elasticity? |
| **Task** | **Task owner:** How to formulate the task to make it more addressable via existing data? How to know what data exists and can be repurposed for my general needs?<br>**Data designer:** How to specify the ideal data schema? | **Managers:** How to foster organizational culture of asking new questions? What are suitable questions for the obtainable data? How to promote culture of innovation and discovery? |
| **Alignment** | **Developers, AI experts:** How to expose data at scale? How to use AI techniques to align data?<br>**Domain experts:** How to integrate domain expertise to inform safe and responsible data alignment? What tools can be used (developed) to support leveraging domain expertise at scale? | **Standard setters:** How to promote data sharing practices such as FAIR principles?<br>**Domain experts:** How to share expertise (e.g., via general-purpose domain ontologies) such that it could be used to support safe and semantically-grounded alignment of data? |
| **Outcomes** | **Managers:** How to get the most out of repurposed data? How to measure the value of repurposing?<br>**Managers, regulators:** How to anticipate the consequences of repurposing? How to mitigate any negative outcomes?<br>**Citizens:** How can the outcomes of (my) personal data be tracked? Is this consistent with the desired uses for my personal data? | **Organizations, Customers:** Are there better products or services due to data repurposing?<br>**Shareholders:** How to generate additional value via data repurposing?<br>**Citizens:** How does data repurposing affect quality of life? |

**Repurposing environment: enablers and constraints.** Data repurposing is affected by the data environment of enablers and constraints. Table 3 illustrates common enablers and constraints in the context of primary components of the framework.

Major enablers and constraints are the knowledge, skills and motivation of those performing repurposing, which can mitigate or worsen the challenges these actors face (see Table 2). Repurposing can require the skills of a wide range of people, organizations and artificial actors (Abraham et al., 2019; Lee et al., 2006). Because repurposing might require bringing together distal sources of data and transforming them under evolving legal and ethical considerations, technical and non-technical expertise might be needed. Proactive support of repurposing involves curating human talent, especially in areas related to understanding data semantics and applying available technologies for identifying, accessing, and transforming data.

For large-scale repurposing to be effective, it is also important to invest in infrastructure. Combining large datasets may require significant computational resources, which will ideally permit scaling based on unforeseen needs. Data lakes and data meshes are solutions for storing high-volume heterogeneous data on distributed database clusters. As Norman (2002) argues, (original) tool designers understand how to use their tools because tool creation is driven by the mental blueprint of the designer. Future users, lacking that mental blueprint, must build their own, potentially flawed, model based on the "combined information" about the tool and objects in the environment the tool interacts with. This also applies to data. To support exposition, for example, it is helpful to consult supplementary resources such as documentation in the form of conceptual models or data dictionaries (Burton-Jones & Meso, 2006; Recker et al., 2021), which increase accessibility and transparency of data and the objects they represent. Administrative agencies can facilitate repurposing by creating well-documented large publicly available data sets, as well as providing infrastructure and training for obtaining and reusing data.

| Table 3. Repurposing enablers and constraints | | | | | |
|---|---|---|---|---|---|
| **Enablers & Constraints** | | **Task** | **Original Data** | **Alignment** | **Outcomes** |
| People | Skills and knowledge | Awareness of sources of data for a task | Ability to understand existing data | Ability to acquire, expose, transform data | Ability to realize analysis into change |
| People | Motivation | Interest in identifying questions to be addressed | Interest in investing effort to use existing data | Interest in developing best available dataset for a task | Motivation to create change, which can meet resistance |
| Resources | Infrastructure | Platforms for task identification (e.g., crowdsourcing) | Storage and meta-data management systems | Large-scale data transformation and computation | Feedback collection platforms |
| Resources | Tools | Automated task identification and conceptualization | Techniques to evaluate data for repurposability | Tools to support and automate alignment | Automated code and system deployment tools |
| Governance | Organizational policies | Guidelines for tasks and cost/effort estimation | Guidelines for original data management | Policies for acquiring, exposing, and transforming data | Guidelines for implementing and evaluating change |
| Governance | Laws / regulations | Determine what tasks might be permitted or prohibited | Determine what data can be collected and how it can be reused | Determine whether data can be legally combined | Determining appropriate actions to take |
| Governance | Norms and ethics | Influence what tasks might be considered appropriate | Influence what data are acceptable to collect and store | Influence whether data are considered representative/biased | Influencing appropriate actions to take |

Policy makers, company executives, economists, ethics experts, journalists, and concerned citizens are increasingly preoccupied with the economic, legal, ethical, and environmental impacts of data repurposing (see Table 2, affected or affecting actors). Hence, repurposing is affected by policies, guidelines and ideas that constantly evolve (Krämer & Shekhar, 2025). The recent thaw in the prohibition on the use of federal funds to study gun violence in the United States (i.e., the Dickey Amendment) illustrates how enablers and constraints can change over time (Weir, 2021). This has permitted researchers to apply for funding through both the Centers for Disease Control and Prevention and the National Institute of Health to study ways to stem the tide of gun violence in the United States (Poltras, 2024). Similarly, in an era where generative AI is stimulating reams of litigation regarding the fair use of data (Bhargava et al., 2025), and the standards of fair use are shifting (Patrick, 2023), there is an evident need for updated data frameworks.

# 4 Applying the Data Repurposing Framework

To illustrate the framework, we apply it to two real-world examples of repurposing, one task-driven and the other data-driven.

## 4.1 Task-driven repurposing addressing environmental challenges via citizen science

We illustrate task-driven repurposing with a case study of repurposing data generated by citizens involved in environmental monitoring (known as citizen or participatory science) (Cooper, 2016; Gura, 2013). This case shows how the Data Repurposing Framework can be used to enhance the benefits and mitigate the challenges of repurposing, as it grew out of a planned original data collection that had to be scrapped due to extraordinary external events.

During the fall of 2019, a master's student of one of the authors planned her thesis on the relationship between light pollution and songbird physiological health in residential settings. The task was well-specified and implied specific data requirements for original data collection that included drawing blood to measure physiological health and using a sky quality meter at night to measure light pollution. However, when the COVID pandemic led to near-universal restrictions in the U.S. in March 2020, the original data collection plan became impossible.

Because the original task was well-defined with an ideal schema comprised of biological markers of bird health from blood samples and sensor measurements of light levels, inherently linked by geographic location, there was a natural starting point for task reconceptualization. This led to considering alternative questions about the nature of human impact on bird health. The outcome was a *redefined task* of understanding the relationship between levels of light and noise pollution and songbird adult survival in residential environments. The team was previously familiar with Neighborhood Nestwatch (citizen generated reports of birds in people's backyards that allows estimation of survival rates of adult songbirds) and the U.S. National Park Service light and noise pollution maps – two data sets with no overlap in content other than being available for the same time periods and geographic coordinates. In this case, the ideal schema of data for the redefined task matched very closely that of the obtainable data, offering confidence that the new question could be answered successfully. Given many years of experience working with environmental datasets, the team knew the data contained in these datasets, the general level of quality of the data and how to address quality shortcomings, how to extract the data, and the constraints on combining and repurposing the data.

Importantly, there were several challenges in repurposing the data. Someone unfamiliar with the data sources would likely not be able to find them, as neither dataset was in public archives. Furthermore, someone unfamiliar with citizen reported data may not appreciate the data quality challenges involved. However, the advisor not only knew about these data sources, but also had collaborations with the researchers stewarding the datasets, who were willing to share them, thereby ensuring accessibility. Researchers who stewarded the Neighborhood Nestwatch dataset restricted the planned reuse by requiring the lead researchers be involved.

To address the new question, the team used 20 years of band-resight data[6] to estimate the annual survival of seven bird species, similar to a procedure used in prior work (e.g., Evans et al., 2015). For each of the nest watch sites in the greater Washington, D.C. area, the team extracted data from the Neighborhood Nestwatch. These data were combined with the second world atlas of artificial night sky brightness, converted to 270-m resolution (Falchi et al., 2016), and with anthropogenic noise data from a map of expected sound pressure levels (Mennitt & Fristrup, 2016). Although measures of impervious surface were extracted from maps with a fine resolution (about 1-m), the sound was sampled at a resolution of about 270m and light at a resolution of about 1777m. All cells were combined so that that all variables matched the coarsest resolution. This resulted in loss of potentially valuable information, which could have been avoided had the light been originally recorded at a finer resolution to make the dataset more elastic. While research questions about survival rates and avian physiological health were not identical, both are relevant to understanding bird conservation, a higher level of abstraction. The primary outcome was increased understanding of the relationship between light and noise pollution and bird survival: noise and light pollution were associate with lower survival of some species, but light was associated with greater survival of others (Pharr et al., 2023).

---

[6] In this approach, unique color band combinations on the legs of birds allowed participants to identify birds as individuals by re-sighting them from a distance.

As the case illustrates, task-based repurposing can be understood and potentially streamlined with the lens of our framework. First, as the framework stresses, actors should be open to redefining tasks in light of available data. This, in turn, necessitates having familiarity with the available sources (or general awareness of where to obtain promising sources) and the ability to form an appropriate model of the ideal data. The extent of the *match* between ideal and obtainable data affects the success or failure of the repurposing effort. Second, many factors may *enable or constrain repurposing*. The case illustrates that some level of domain expertise and technical prowess is needed to handle acquisition, exposition and transformation. As the case shows, data often need to be remixed in non-trivial ways (augmentation with seemingly unrelated sources, mashups with spatial and temporal data, finding common connectors and granularity levels). Broadly, this means that an *environment* in which people having the requisite skills can greatly facilitate repurposing. Third, the case shows the impact of original data management on repurposing, such as in shaping the use-agnostic dimensions of transparency and elasticity. Finally, it is notable that the *project outcomes* not only included answering the immediate question, but could also lead to delayed outcomes, such as follow-up questions about why some animals appear to benefit, while others suffer, from anthropogenic activities. In the longer term, this data repurposing might contribute to policy outcomes such as efforts to mitigate noise and light pollution in built environments.

## 4.2 Data-driven repurposing in the healthcare context

We further illustrate data-driven repurposing with a case study of repurposing an existing dataset. The Hospital Discharge Dataset, managed by the Florida Agency for Healthcare Administration (AHCA), provides a census of patients admitted to acute care facilities in Florida since 1988. These data have been extensively repurposed for academic research (e.g., Burke et al., 2007; Greenwood & Agarwal, 2015), as well as for policy purposes.[7] Even so, we are unaware of a systematic evaluation of the suitability of the AHCA data for repurposing. By applying the Data Repurposing Framework, we show how we can gain insights not reflected in the publications using this data set and identify challenges when dealing with data.

---

[7] AHCA data have been repurposed to support many policy decisions, such as Medicaid data visualization (https://ahca.myflorida.com/medicaid/medicaid-finance-and-analytics/medicaid-data-analytics/medicaid-data-visualization-series), consumer decision making (https://ahca.myflorida.com/agency-administration/florida-center-for-health-information-and-transparency/office-of-data-dissemination-and-transparency), and analysis for disease tracking and related uses (https://www.flhealthcharts.gov/charts/HealthResourcesAvailability/default.aspx).

We first assess the data's natural suitability for repurposing, focusing on the framework's dimensions of accessibility, transparency, and elasticity. The AHCA dataset was created with the expressed directive of being Florida's "chief health policy and planning entity." This means the dataset was intended to support data reuse and repurposing for policy setting. The AHCA dataset is the product of well-articulated data management practices governed by statutory authority, thereby inviting attention to a broad class of questions relating to public and population health, including costs, patient care outcomes, administrative benchmarking, and monitoring trends in the population. As with other administrative datasets, the AHCA dataset is designed to be accessible to support new tasks. The fields of the dataset are described in a well-organized data dictionary accessible online.[8] Potential users can select the data they wish and submit a request for approval and billing, after which the user is sent a link with credentials to access it.

The ACHA dataset is also transparent as it can be understood by consulting a data dictionary. Additional documentation is available to facilitate use. Still, some transparency challenges remain, as the use of AHCA requires the analyst to have basic training in the use of medical data (e.g., understanding what DRGs and ICD-9/10 codes are). There is an opportunity to make this dataset even more transparent by providing additional resources to support those less familiar with the medical and insurance domains.

The data have substantial elasticity, in part due to the level of granularity at which they are reported. Discrete hospital admissions are tracked at a quarterly level. The data can be linked with other datasets using geographic information (e.g. from the census) or hospital level information (e.g., via Hospital Compare) because those are stable and visible, providing further elasticity. At the same time, the dataset has elasticity challenges. For example, there are no stable patient identifiers, making it impossible to track patient outcomes over time. In addition, users must pay to access data, thereby limiting repurposing to those with adequate resources. Nonetheless, because these data have been repurposed so extensively, this demonstrates substantial elasticity.

The second analysis deals with data-task alignment. We can examine the extent to which the AHCA can be acquired, exposed, and transformed to be suitable for new tasks. Here, the strengths and weaknesses of these data are apparent. For example, the presence of stable physician and hospital identifiers permits researchers to augment the AHCA data by combining it with datasets containing location or physician specific information, such as a localized treatment or time varying physician data (e.g., Greenwood & Agarwal, 2015; Lu & Rui, 2018). Alternatively, the lack of information beyond the attending physician and surgeon, and the lack of stable patient identifiers, precludes work related to the broader patient care team (e.g. nurses, support staff) or to specific patients over time.

While the data's exposition is transparent, being well documented by the data provider, its suitability for transformation is contingent on the task (i.e., the corpus of patients the researcher wishes to study). While the data can be filtered based on factors like patient sex or race to understand medical outcomes, or the effects of external developments like the emergence of medical knowledge (e.g., Greenwood et al., 2017), the infrequent admittance of patients with psychiatric conditions, for example, precludes their use on a large scale because any empirical design will lack statistical power (i.e., these patients are largely unobserved). Moreover, while datasets like Hospital Compare can easily be combined with the AHCA data to understand general patient experiences (e.g., Grimsley et al., 2023), patient privacy rules preclude the researcher from investigating individual experience, which is idiosyncratic to a single patient.

---

[8] The data are described here: https://quality.healthfinder.fl.gov/Researchers/Order-Data/.

As we see from this case, the framework enables us to better *understand* and *facilitate data-driven* repurposing. First, our framework allows the assessment of the strengths and weaknesses of a particular data set for repurposing. Second, the framework can also guide more effective repurposing by pointing out specific challenges with a data set that can be given greater attention when repurposing. Finally, our framework can also inform the future updates to the data set to make it more amenable for a purpose and, by making it more accessible, transparent, and elastic.

### 4.3 Insights from the Cases

Considering both cases, it is clear that the framework brings into focus the key aspects of the varied and complex practice of repurposing. It helps to understand both task-driven and data-driven repurposing, thereby unifying disparate practices. It also responds to repeated calls on explicating challenges in repurposing, such as how to understand the role of original data management or evaluate data quality in a use-agnostic way (Sadiq et al., 2022; Strengholt, 2020; Zhang et al., 2019). Finally, it suggests aspects of repurposing that are critical to success, but poorly understood, including: ability to articulate an ideal data schema, closeness between ideal and obtainable data, data set familiarity, willingness to redefine the task, capability to envision new uses for data, understanding the limitations of specific data sets, understanding how to address these limitations, and understanding how to expose data to ensure transparency.

## 5 Research opportunities in data management for repurposing

The Data Repurposing Framework suggests important opportunities for future data management research. We follow the structure of the framework in highlighting these opportunities.

### 5.1 Original data management and properties of original data

The Data Repurposing Framework introduces three use-agnostic properties of data: accessibility, transparency, and elasticity. These dimensions are shaped by factors such as technologies, the availability of metadata, and the structure of the data. Research is needed to better understand the antecedents and consequences of each of these dimensions. For example, how do traditional and emerging methods for collecting and storing data affect the use-agnostic dimensions?

The framework broadens the concept of data independence to include conceptual independence. Not all data can be easily decoupled from its original schema. For example, in crowdsourcing contexts original data collection driven by fixed predefined categories (e.g., biological species—a type of schema) may be undertaken to ensure data satisfy quality requirements for predetermined needs. However, this makes the resulting data less amenable to unanticipated insights (Lukyanenko, Parsons, Wiersma, et al., 2019). In contrast, data collected as free-form attributes and categories describing observed objects as they are understood by contributors yield more discoveries and can be reused in more ways. Considering the above crowdsourcing example, conceptual independence is extended to include the extent to which data are stored independent of a fixed set of categories (Parsons & Wand, 2000).

Conceptual independence stands to mitigate limitations arising from interpreting data from the perspective of predetermined original uses, thereby facilitating use for unanticipated tasks. At the same time, some description of an object is unavoidable and questions remain as to how much description is enough and what form it should take to promote conceptual independence. Insights on these questions can be drawn from research on classification and data-interpretation biases (Murphy, 2004), as well as from cognitive and ontological theories that suggest alternative approaches to data collection and storage (Parsons & Wand, 2000).

Repurposing adds a new consideration in that overemphasizing data independence may come at the expense of other repurposability dimensions, such as transparency. For example, original data may be enriched with additional semantic hierarchies (e.g., spatial hierarchies such as city-state-country) that, while framing these data and reducing data independence, can help with its interpretation (Sleeper, 2020). Research is needed to investigate the effect of data independence on other dimensions of repurposing and to mitigate negative consequences.

As the framework shows, original data management directly shapes the repurposing dimensions. An open question is how to balance the need of fitting data to the original purpose with maximizing its repurposing potential. For example, if data must be anonymized for privacy, this will limit its repurposability. Conversely, collecting the most specific data possible for the original task requirement will increase its elasticity. Hence, there is an opportunity for devising methods that simultaneously promote fitness for the original purpose and data repurposability.

Finally, future studies can investigate the relationship between the traditional and use-agnostic properties of data. What is accurate and complete for one task may be inaccurate or incomplete for another, but this does not imply that these original data properties are irrelevant for repurposing. For example, if the original need is to record sales by day and data is stored as the correct day but an incorrect hour, the data would be accurate for the original and other purposes (e.g., analyze sales by quarter). However, the inaccuracy would constrain other tasks (e.g., tracking sales by hour) and raises concerns about the integrity of the data collection process.

New concepts, including data products and data commodities, are becoming popular in contexts such as machine learning, data science, news and finance. These terms denote data that is well-documented and free of errors, and thus amenable to a variety of uses (Plotnikovs, 2024). One can argue the AHCA dataset exemplifies these concepts, although our analysis shows it has room for improvement. As the industry of data products expands, a variety of questions emerge. One is what are the metrics by which the quality of data products can be gauged? Our framework provides guidance on this as it suggests a combination of use agnostic (accessibility, transparency and elasticity) and fitness for use dimensions (accuracy, completeness). Future studies can create an index for data products which would account for these quality dimensions, potentially considering different settings. Another question is how to derive the economic value of data products from data management practices, such that the economic impact of data management can be better understood and appreciated.

## 5.2 Repurposing Activities

As the Data Repurposing Framework shows, repurposing includes activities different from those of original data management, each presenting research opportunities.

**Feasibility assessment**. At the early stages, one might assume that any data project involves collecting data from scratch. However, with the growth of data it is more reasonable to start with the presumption that existing data could satisfy the needs of a task. Yet existing data management frameworks (Chua et al., 2022; DAMA et al., 2024) lack a feasibility analysis stage for data repurposing. The problem is not trivial, as the value of repurposing may only be evident after costly data acquisition and transformation (Grande et al., 2020; Parmiggiani et al., 2022). Hence, an important research challenge is estimating the cost, risk and value of repurposing versus collecting data from scratch. Accumulating additional empirical evidence of the benefits and challenges of repurposing (e.g., Plotnikovs, 2024) is a valuable aspect of data management for repurposing.

**(Re-)Conceptualization.** There is a large body of work on conceptualizing data for specific tasks (for reviews, see Chua et al., 2022; Recker et al., 2021; Storey et al., 2023; Wand & Weber, 2002). Traditional approaches to conceptualization assume the resulting data will align well with the requirements of the task, but this is often not the case in repurposing. Data-driven repurposing starts with the obtainable data, even when it is far from ideal. For example, consider research using the size of signatures of CEOs on audit documents as a proxy for narcissism, as in Chou et al. (2021). A task-driven approach can be used to conceptualize an ideal dataset for the task (e.g., a brain scan of current CEOs to gauge their level of narcissism), which serves as a baseline against which to evaluate and, if necessary, augment available data. These issues are beyond the scope of traditional data management. As a result, they have received little attention from scholars, and many questions remain, including how to estimate and quantify the difference between ideal and actual data, how to appropriately use ideal data models that do not correspond to the actual data, how to connect multiple related conceptualizations, and how to design conceptual modeling methods and grammars to support reconceptualization for repurposing? Any design efforts addressing these issues should be supported by behavioral and theoretical studies of the impact of proposed solutions on the range of roles involved in repurposing.

**Acquisition.** The expansion of data sources creates new challenges in identifying, monitoring, and searching for appropriate data. Finding relevant data has been studied extensively in information retrieval, web search and database contexts (e.g., Arazy & Woo, 2007). A problem in repurposing is that data become valuable for a new task only when aligned with that task. The potential of a specific dataset for a repurposing task might not be clear until costly transformations are undertaken (Grande et al., 2020; Parmiggiani et al., 2022). Here, important research questions are: How can data be screened to identify sources of high potential? And how can these sources be assessed for repurposability without engaging in full-scale transformations? A related challenge is determining the cost of obtaining and using data early in the process. Insights for addressing these issues can come from analytical modeling to determine optimal data purchasing decisions given the attributes of the data (e.g., Liu et al., 2022).

The search for suitable data for repurposing is challenging in part because drawing new insights often requires imagination and serendipity. Open challenges include how to better identify relevant data for the task and, conversely, how to identify fruitful tasks given the available data sources. One underexplored opportunity here is crowdsourcing, which demonstrates the impressive ability of domain non-experts to think outside the box (Brabham, 2009; Vermicelli et al., 2021). Ideas for using crowdsourcing can be drawn from research on crowdsourcing design to promote discovery (e.g., Cooper, 2016; Lukyanenko, Parsons, Wiersma, et al., 2019), as well as work on crowd innovation (Amabile, 2019).

**Exposition.** Transparency on how original data were managed (e.g., conceptualized, collected) is necessary for drawing valid insights from repurposing, as well as for using data safely and responsibly. Common ways to facilitate transparency are to publish data openly in a standard format (such as RDF) or provide comprehensive metadata that give context to the data. In this vein, organizations increasingly publish data collected from public and private sources, as with the AHCA dataset. To support these efforts, open data formats and other techniques, such as knowledge graphs and domain ontologies, are used. To expand the scale of these activities, research on semantic annotation of datasets, such as automated annotation using artificial intelligence, manual annotation using crowdsourcing, or hybrid approaches, is needed. Likewise, domain ontologies are laborious to create and validate (Burton-Jones et al., 2005; McDaniel & Storey, 2019). In addition to research on ontology generation and validation, there is an opportunity to better use general, or upper-level, ontologies (e.g., Unified Foundational Ontology, Bunge, Dolce) (Guizzardi, 2005) as a basis for developing and validating domain specific ontologies to increase transparency. As there has been a proliferation of general ontologies, systematic investigation of their potential to support data exposition for repurposing is needed.

Often, little is known about how to identify why and under what circumstances data were created. While motivation is often clear in the case of structured organizational data, this is not the case for unstructured social media posts or amateur journalism, where different reasons for capturing data may result in different analyses and conclusions. Few methods exist for systematically capturing the intent, motivation, assumptions, and situational factors of data collection. Correspondingly, there is no systematic way to extract intent. Similarly, legal, ethical, and privacy concerns are heightened when using data from online sources, necessitating more research on detecting disinformation and protecting the privacy and intellectual property of content creators. There is an opportunity for data management researchers to work with scholars from other disciplines that study citizen science, open journalism, and social media, where motivations of the creators has been known to affect conclusions drawn from such data (Tong & Zuo, 2021). Furthermore, data management can benefit from advances in AI, such as large language models (LLMs) (Vaswani et al., 2017). For example, LLMs can be used to capture sentiment in social media data to help infer the underlying assumptions of the creators.

**Transformation**. How original data is transformed affects how well it can be aligned with a new task. Many techniques exist for combining data in the context of database schema matching and semantic interoperability (Batini et al., 1986; Litwin et al., 1990; Sheth & Larson, 1990), but repurposing adds new challenges. As earlier work on schema matching and semantic interoperability originated in structured database settings (Batini et al., 1986), developing and improving techniques for heterogeneous, semi-structured and unstructured data integration remains an important opportunity. As LLMs increasingly repurpose multimodal data (text, videos, sounds, images), integrating data of different modalities is a research frontier.

Traditional research on semantic interoperability assumed the alignment process is user-driven and hence, transparent to the user (Friedman et al., 1999; Sheth & Larson, 1990). This assumption no longer holds as transformations are often automated, including the case of synthetic AI-generated data (Guo & Chen, 2024) or automated machine learning (Larsen & Becker, 2020). Danger exists when transformations are opaque and it is unclear what decisions were taken when data were transformed or what information was lost in the process. Repurposing data without understanding how it has been transformed can lead to inappropriate conclusions and actions. Despite the rise in these practices, more research is needed on the impact of the transparency of data transformations on trust in, and adoption of, data transformation tools. Furthermore, there is a design-oriented opportunity to increase the transparency of data transformation tools.

One data management challenge is how to manage data quality tradeoffs. As illustrated in the case of AHCA, improving upon one dimension of data quality (e.g., increasing completeness by recording personal identifiers) can come at the expense of other considerations (e.g., loss of privacy). Managing the tradeoff between completeness and privacy is particularly important in repurposing data about humans, as it can limit the extent to which a dataset can be aligned with a particular task. The data management community can explore these and other issues by collaborating with design and behavioral privacy researchers.

Another transformation challenge is combining data from multiple heterogeneous sources. Traditional data integration focuses on matching data from well-structured sources (e.g., database schema mapping) (e.g., Batini et al., 1986). However, as the Neighborhood Nestwatch case shows, data from different domains and of varying levels of quality (e.g., granularity) are being combined. For example, light and noise pollution data had to be matched on geo-coordinates with bird sighting data from backyards, with datasets varying in their level of granularity. Data management researchers can contribute to research on semantics of heterogeneous and multimedia sources, management of open data, and enrichment of stochastic models (the basis of tools such as ChatGPT) with domain knowledge, through direct and indirect human-in-the-loop approaches (e.g., crowdsourcing, conceptual modeling, ontologies).

## 5.3 Enablers and constraints

Repurposing does not concern data alone, as data interact with the broader environment. Research is needed on the impact of data environment objects on repurposing. Traditionally, this area, commonly understood as context, has been overlooked in data management, and this is even more acute for repurposing. Our framework conceptualizes the environment in terms of enablers and constraints of repurposing. As there could be many enablers and constraints, a basic objective is to identify the systems and objects having the most impact on repurposing.

Many repurposing challenges stem from the ongoing evolution of data storage and use. First, many new storage and access systems disrupt data management, including APIs, data lakes, data meshes, and new data access tools (Inmon & Srivastava, 2023; Strengholt, 2020). APIs, for example, are a key tool for data exchange, directly affecting how external agents interact with data. Yet, little methodological guidance exists for designing and modeling APIs to facilitate repurposing. Traditional data management assumes structured relational data (Chua et al., 2022; Codd, 1970). However, much repurposing occurs from NoSQL databases, data lakes and other scalable and distributed systems (Atzeni et al., 2020; Giebler et al., 2019). Furthermore, repurposing often involves combining data from different sources using different formats. It is an open question how to model extremely heterogeneous, multi-modal data effectively. A well-organized data lake prevents it from becoming a dumping ground for all kinds of data (a "data swamp") but imposing too much or the wrong structure can stifle repurposing. Insights on these questions might be obtained from work on modeling individual objects (Lukyanenko, Parsons, & Samuel, 2019; Parsons & Wand, 2000), flexible data design (Lukyanenko et al., 2024), (data) modularity (Norman, 2002), and basic level categories (Castellanos et al., 2020).

Second, the rise of AI systems is fueled by their ability to ingest massive amounts of data for training. This has led to new data management challenges for both data owners and the organizations behind these systems. Our framework can guide relevant research from both perspectives. On the one hand, the quality of data used to train AI affects model performance. Consider prompts to an LLM about some medical condition. A model trained on vetted sources would be more appropriate and safer to use than one trained on grey literature. As the framework makes clear, understanding the sources of data and how they are generated (via exposition) is a prerequisite for data-task alignment. Hence, more work is needed on how to expose the provenance of data before it is used in AI. This is contrary to the "kitchen sink" approach sometimes taken in practice, where as much data as possible is ingested to find useful patterns.

Another area needing research covers policy and regulatory strategies related to fostering and regulating repurposing. One particular question is the implications of using training data from public sources, and design-oriented questions of how to create more regulation-compliant extraction methods. Furthermore, it is notable that OpenAI, the developer of ChatGPT, has been sued for copyright violations.[9] This illustrates the legal and ethical enablers and constraints around data ownership that should be considered in planning repurposing activities.

Another research opportunity is fostering a work culture that makes repurposing successful. For all repurposing activities to be successful, there need to be sufficient capabilities. Hence, related to exposition Alaimo and Kallinikos (2024, p. 66) suggest: "[r]eusing and repurposed data requires reimagining their roles in building their organizational capabilities to fuse such operations with established organizational and knowledge functions and practices." Reflecting on issues related to getting most out data in the context of Microsoft, Plotnikovs (2024) points to the surprising deficit of data culture in that organization. The academic community can do more to understand such issues and improve existing practices. Culture is a multidimensional construct in organizational and social sciences (Homburg & Pflesser, 2000) and is in need of better formalization and measurement. A key question is how different dimensions of data culture affect repurposing by enabling or constraining different repurposing activities.

The data environment also includes organizations, governments and other bodies that seek to regulate and govern repurposing. Rampant data repurposing raises serious legal, ethical and privacy challenges. Our framework suggests considering original data management as a starting point for approaching these issues. Indeed, how one conceptualizes, collects, curates, consumes, and controls original data affects its repurposability and compliance with legal, ethical, and privacy norms. For example, differential privacy – introducing noise or purposefully altering data when data is curated to hide sensitive data, but still preserving the signal (Menon et al., 2005) – can both enable and constrain repurposing and its impacts need to be better understood.

To answer certain questions in a truly innovative way, one must think outside the box and find sources that on the surface have nothing to do with each other. To pave the way for drawing greater value from data, it is necessary to design policies and promote a data management culture that encourages data serendipity, supported by research to investigate the effects of such practices, especially on organizational and societal outcomes.

---

[9] https://www.nytimes.com/2023/12/27/business/media/new-york-times-open-ai-microsoft-lawsuit.html

Finally, data repurposing raises questions about fairness and equity. Access to relevant data is a prerequisite for repurposing, yet data are not equally available for all (e.g., the cost to access AHCA data). There is a need for studies to investigate and find solutions to mitigate data inequity and reduce barriers to data sharing and data transparency. Kramer and Shekhar (2025) note significant advantages when platforms repurpose their data to spawn new businesses. Considering policy options for ensuring these benefits are equitably distributed in the market while promoting innovation is an open research opportunity.

## 5.4 Repurposing Actors

As Table 2 shows, there a numerous actors involved in data repurposing. Some are involved directly, such as those who formulate the tasks and perform data transformations. To this point, only the practitioner literature has addressed the challenges of these direct actors, primarily from the perspective of the tools and techniques to follow in handling data (Plotnikovs, 2024; Strengholt, 2020). There is little work on understanding the role of the actors involved in repurposing, especially the challenges and needs of those indirectly involved and affected by repurposing. Theoretical insights and empirical evidence can be valuable, especially on topics such as the psychology of actors, organizational capabilities that foster repurposing, and ethical and legal foundations of repurposing. Table 2 lists numerous challenges faced by direct and indirect actors, which shape specific research opportunities to address these challenges.

## 5.5 Repurposing outcomes

A final major research opportunity lies in understanding the outcomes of repurposing. The outcomes include the production of another data source, the *repurposed data*. This differs significantly from the original data due to extensive transformations and enrichments. It can also become an input for further repurposing, potentially by another team, which creates challenges. Traditionally, when datasets are created, they are supported by documentation such as conceptual models or data dictionaries. These typically capture the form and structure of domains (Burton-Jones & Weber, 2014) but not how data was transformed. Furthermore, many practices are new, such as the generation of synthetic data by AI for repurposing as training data for new tasks. Given the impact of transformation on repurposed data, an important opportunity is developing approaches for documenting data transformations to support replicability, transparency, explainability, and further repurposing. This work can benefit from collaboration with the machine learning community, which studies ways to document feature engineering and data transformations (Duboue, 2020).

Work on the outcomes of repurposing should also consider how original and repurposing data management activities and properties of data affect organizational or societal outcomes. Related studies focus on the benefits of open data sharing and data transparency (Chui et al., 2014). In conducting this work, it is important to relate the properties of data (such as its accessibility, transparency, and elasticity) to both immediate and distal outcomes.

A further research opportunity is to investigate the benefits and negative impacts of repurposing. To date, much evidence in support of these outcomes is anecdotal. For example, for organizations that rely extensively on repurposing, the costs of sourcing, transforming and using data can be as high as 25% of total IT spending (Grande et al., 2020). It is unclear whether these costs generalize to all organizations or are contingent on factors such as skills, resources, methods, and approaches used for data management.

The success of data repurposing needs to be considered against its costs, including needed skills and resources, infrastructure, and training, as well as detrimental consequences for individuals and society (e.g., erosion of privacy). An overlooked consequence of data repurposing is its environmental impact. Further research is needed to measure and mitigate the environmental consequences of data repurposing versus original data use. More broadly, with the potential of data to affect long term performance of organizations and economies in often unforeseen ways, more systematic and comprehensive cost-benefit and risk analysis of data repurposing is needed.

Recall that the challenges of indirect outcomes are amplified for repurposing. This creates new research opportunities. For example, what is the range or scope of unpredictable outcomes of repurposing? How can uncertainty be reduced? What are the most critical issues (e.g., privacy, security, reputation, consent, or intellectual property) to be better anticipated? As our framework highlights the importance of original data management to repurposing, an important question is how some uncertain outcomes can be anticipated when data is collected for primary purposes.

Table 4 summarizes the research agenda arising from the Data Repurposing Framework. This agenda seeks to address the open questions arising from prior work (Table 1). Specifically: (1) questions about the characteristics of data that enhance repurposability require understanding the relationship between traditional and use-agnostic properties of data; (2) questions about technological solutions to repurposing data require understanding the benefits and limitations of doing so in semantically meaningful ways; (3) questions about assessing the repurposability of data in advance require understanding both ideal and available schemas, and how to determine the extent to which the gap between them can be bridged; and (4) the implications of data repurposing require a better understanding of both short-term and long-term benefits and costs.

## 6  Conclusions

Data repurposing is pervasive, but it introduces challenges and risks. We develop a Data Repurposing Framework focusing on the unique challenges and opportunities of repurposing data and identify important research questions for doing so meaningfully, effectively, and appropriately. We also call for research that broadens the scope of data management and builds bridges between data management researchers and scholars in other disciplines.

| Table 4: Research agenda for data repurposing | | |
|---|---|---|
| Research Area | Issues | Research Opportunities |

| Original data management and use-agnostic properties of original data | <ul><li>Data collection/storage decisions affect use-agnostic properties</li><li>Data independence</li><li>Relationship among data quality dimensions</li></ul> | <ul><li>How to manage data to maximize repurposability while meeting the needs of known use</li><li>Understanding the relationship between traditional and use-agnostic properties of data</li><li>How data independence affects repurposing</li></ul> |
|---|---|---|
| Repurposing activities | <ul><li>Match between task and data is key to success</li><li>Data and task may need to be re-conceptualized during repurposing</li><li>Acquiring the data to support tasks</li><li>Understanding the data being repurposed</li><li>Undertaking appropriate transformations</li></ul> | <ul><li>How to foster discovery of relevant data</li><li>How to estimate the match between task and data</li><li>How to measure and manage the gap between ideal and available data</li><li>How to assess the potential of obtainable data at reasonable cost</li><li>How to expose and transform data, including using AI</li></ul> |
| Repurposing enablers and constraints | <ul><li>The environment in which repurposing occurs affects activities and outcomes</li><li>Evolving technological landscape creates new challenges and opportunities</li></ul> | <ul><li>How to effectively use the technological enablers of repurposing including AI</li><li>How to repurpose effectively within legal, ethical and moral frameworks</li></ul> |
| Actors | <ul><li>Challenges as listed in Table 2, Challenges for Actors Involved in Data Repurposing</li><li>Recognizing the under-appreciated indirect actors involved and affected by repurposing</li></ul> | <ul><li>Developing tools and techniques to support repurposing actors</li><li>Advancing theoretical foundations of challenges of direct and indirect actors</li><li>Understanding the actors and their challenges</li></ul> |
| Repurposing outcomes | <ul><li>Repurposed data may be repurposed again</li><li>Outcomes may be indirect and delayed</li><li>Data fusion may offer unique competitive advantages</li></ul> | <ul><li>How to further repurpose repurposed data</li><li>Understanding how use-agnostic properties of data affect repurposing outcomes</li><li>Better understanding the long-term benefits and costs of repurposing</li><li>Regulating unequal benefit distribution due to data repurposing</li></ul> |

## Acknowledgements

The authors wish to than the Senior Editor, Andrew Burton-Jones, the Associate Editor, Sumit Sarkar, and the anonymous reviewers for their constructive feedback on prior versions of the paper. We also thank participants in the McIntire School of Commerce brownbag seminar, especially its organizer, Professor Peter Gray, for their feedback, and students in the UVA MSMIT program for the valuable insights on the practice of data management for repurposing. We are grateful to Ron Weber for helpful discussions on the nature of data and digital.

**Funding.** This research was supported in part by funding from the Social Sciences and Humanities Research Council of Canada to Jeffrey Parsons [Grant 435-2020-0402].

## Biographies

**Jeffrey Parsons** [0000-0002-4819-2801] is University Research Professor and Professor of Information Systems in the Faculty of Business Administration at Memorial University of Newfoundland in Canada. His research focuses on how to better represent human conceptualizations of the world in data. His work on this and related topics has been published in several disciplines. Jeff's research has been recognized in several ways, including *MISQ* paper of the year (2019), AIS Senior Scholars Paper Award (2020), and the INFORMS Design Science Award (2014). He is a Fellow of the Association for Information Systems, Distinguished Research Fellow from TU Dresden, Schoeller Senior Fellow, and an ER Fellow. He has served as a senior editor at the MIS Quarterly and is current a senior editor at Information Systems Research.

**Roman Lukyanenko** [0000-0001-8125-5918] is an associate professor at the McIntire School of Commerce, University of Virginia. His research interests include data management and research methods (validity and artificial intelligence in literature reviews). Roman actively develops ideas, tools, and methods to improve data management and research practices. These solutions received major awards, including INFORMS Design Science Award, Governor General of Canada Gold Medal, Hebert A. Simon Design Science Award. Roman's research has appeared in Nature, MIS Quarterly, Information Systems Research, ACM Computing Surveys. His 2019 paper on quality of crowdsourced data received the Best Paper Award at MIS Quarterly.

**Brad N. Greenwood** [0000-0002-0772-7814] is the Dean's Distinguished Professor of Business at George Mason University's Costello College of Business. He has also served on the faculty at University of Minnesota's Carlson School of Management, Temple University's Fox School of Business and the University of Maryland's Smith School of Business. Dr. Greenwood's research examines the intended and unintended consequence of innovation, and how access to the resulting information affects welfare at the interface between business, technology, and social issues; notably in the contexts of healthcare and entrepreneurship. He is currently an Associate Editor at Management Science and a Senior Editor at MIS Quarterly. His corporate experience includes nearly eight years as a deputy project manager and analyst for CACI International. He received his Bachelor's degree in Information Technology and Management Information Systems from Rensselaer Polytechnic Institute. Dr. Greenwood also received his MBA from the University of Notre Dame; his Master's of IT from Virginia Polytechnic Institute; and his PhD from the University of Maryland, College Park. His degrees in law are from George Mason University's Antonin Scalia Law School.

**Caren B. Cooper** [0000-0001-6263-8892] is a professor of public science in the Department of Forestry & Environmental Resources in the College of Natural Resources at North Carolina State University. Her research spans environmental sciences, social sciences, and humanities. She uses participatory research methods to study a range to topics related to the environment, particularly birds, soundscapes, and water quality. Her social science research focuses on understanding and improving the participatory sciences related to issues of data quality, participant diversity, and participant learning. In the humanities, she led the development of the Data Ethics Toolkit for Participatory Sciences. She is author of *Citizen Science: How Ordinary People Are Changing the Face of Discovery* and co-author of *The Field Guide to Citizen Science: How You Can Contribute to Scientific Research and Make a Difference*.